\documentclass[epj]{svjour}
\usepackage{graphicx,epsf,dsfont,amssymb}

\usepackage{bm}% bold math
\usepackage{amsmath} % \pmb bold special characters
 \RequirePackage{color}
 
\def\bsigma{{\pmb{\sigma}}}

\def\nbOne{\pmb{%\mathbb 
1}}
\def\be{\begin{equation}}
\def\ee{\end{equation}}
\def\bey{\begin{eqnarray}}
\def\eey{\end{eqnarray}}
\def\matW{{W}}
\def\mats{s}
\def\nadermat{{{\cal D}}} 
\def\vac#1{{\bf{#1}}}
\def\DeltaSO{\Delta_{{\rm SO}}}
\def\lambdaSO{\lambda_{{\rm R}}}

\begin{document}
\date{\today}

\title{Gauge transformations of Spin-Orbit interactions in graphene}
\author{Bertrand Berche\inst{1,}\inst{2} 
	\and Nelson Bol\' ivar\inst{1,}\inst{2,}\inst{3}
	\and Alexander L\'opez\inst{4,}\inst{5}
	\and Ernesto Medina\inst{2,}\inst{1,}\inst{3}
	}
\institute{Groupe de Physique Statistique, Institut Jean Lamour, Universit\'e de Lorraine, 54506 Vandoeuvre-les-Nancy, France
	\and Centro de F\'isica, Instituto Venezolano de Investigaciones Cient\'ificas, 21827, Caracas, 1020 A, Venezuela
	\and Escuela de F\'isica, Facultad de Ciencias, Universidad Central de Venezuela, 1040 Caracas, Venezuela
	\and Institute for Theoretical Physics, University of Regensburg, D-93040 Regensburg, Germany
	\and Universidad de Investigaci\'on de Tecnolog\' ia Experimental YACHAY
	Ciudad del Conocimiento, San Miguel de Urcuqui, Ecuador-100119
}

%----------------------------------------------------------------------------
%                             ABSTRACT
%----------------------------------------------------------------------------
\abstract{
Inclusion of spin-dependent interactions in graphene in the vicinity of the Dirac points can be posed in terms of non-Abelian gauge potentials.
Such gauge potentials being surrogates of physical electric fields and material parameters, only enjoy a limited
gauge freedom.  A general gauge transformation thus in general changes the physical model. We argue that this property can be useful
in connecting reference physical situations, such as free particle or Rashba interactions to non-trivial physical Hamiltonians with a new set
of spin-orbit interactions, albeit constrained to being isoenergetic. We analyse different combinations of spin-orbit interactions in the case of monolayer graphene and show how they are related by means of selected non-Abelian gauge transformations.
% end of PACS codes
\\
{\bf keywords}: graphene, spin-orbit interaction, non-Abelian gauge theory, gauge transformation
} %end of abstract
\maketitle

\section{Introduction\label{sec1}}
The spin orbit interaction (SOI) is rapidly becoming a tool for spin manipulation in material systems of high technological importance such as non-centrosymmetric semiconductors and of potential relevance such as graphene-like compounds\cite{Nowack,Kouwenhoven}. It is also
a crucial ingredient in the creation of surface chiral transport states\cite{Kane,Bernevig} of topological insulators where
the bulk gap is created by the SOI.  The interaction of relativistic origin, couples the spin to an electric field whose source can be internal, due to the atomic cores, or externally applied via voltage gates. The spin-orbit coupling enters the Hamiltonian in two variants, i) the Rashba type coupling that involves an external electric field via structural inversion asymmetry\cite{Winkler} produced for example by a gate and ii) the intrinsic spin-orbit coupling, produced by the inversion asymmetry of the bulk material as in zinc-blende crystals or purely by the core penetration of the atomic wave functions\cite{Winkler}. Analogous interactions are defined for two dimensional materials such as graphene, where Rashba type interactions
are a result of a combination of the atomic SOI and the Stark effect due to an external field, while the intrinsic coupling is only a function
of the lattice structure and the atomic SO coupling. 

The spin-orbit interaction might appear as a peculiar interaction to describe in a Hamiltonian formulation since it depends on
momentum, so in a sense it is part of the complete description of a kinetic energy operator. In this sense there is a
mathematical connection to gauge theory, where one can reinterpret the SOI as a gauge vector and gain insight
into the problem by way of the connection to well established formalisms in QED and  gravity\cite{Vozmediano}.

As a direct illustration let us take the Pauli SO coupling where there are no external magnetic fields.
\begin{equation}
H=\frac{{\mathbf p}^2}{2m}-\frac{e{\boldsymbol s}\cdot\left({\bf E}\times 
{\mathbf p}\right)}{2m^2c^2},
\label{PauliFirst}
\end{equation}
where ${\mathbf p}$ is the momentum operator, ${\boldsymbol s}$ is the electron spin, ${\bf E}$ is the electric field
$m$ is the electron mass and $c$ is the velocity of light.
It is immediately suggestive to complete the square and construct a Hamiltonian that is reminiscent
of a minimal coupling in electromagnetism.
\begin{eqnarray}
H&=&\frac{1}{2m}\left ({\mathbf p}-\frac{{\bf W}^a s^a}{2}\right )^2
\nonumber \\
&&-\frac{\hbar^2}{8m}{W}_i^a{W}_i^a
\label{PauliGauge}
\end{eqnarray}
where ${W}_i^a=(e/ 2mc^2)\epsilon_{aji}{ E}_j$ can be seen as a new $SU(2)$ gauge field
and the definition of a minimal coupling to the spin is done by the substitution $p_i\to p_i- W^a_i\frac{s_a}2$. This is equivalent  
to defining the covariant derivative
${\nadermat}_i=\partial_i -\frac i\hbar W^{a}_{i}\frac{\mats_a}{2}$.  Here, the $SU(2)$ gauge field is 
$W^a_i$ and the generators of $SU(2)$ algebra, given in terms 
of the spin components $\mats_a$ where $\mats_a=\hbar\sigma_a/2$ and $\sigma_a$ is a Pauli matrix.
In the expression $W^{a}_i{\mats_a}$, the spin index $a$ is contracted and the matrix structure of the non-Abelian
gauge field is apparent. The space index $i$ runs 
over $i=1,2$ in standard applications to two-dimensional
non-relativistic electron gas (2DEG).  {It is important to note that in this
formulation the gauge field is anchored to physical quantities such as material parameters and
the electric field\cite{Vozmediano}.}

Such a gauge point of view, reformulating in general terms the spin-orbit interaction as 
a $SU(2)\times U(1)$ gauge theory in the non-relativistic case, has been known for some 
time~\cite{Goldhaber,Mineev,Frohlich}. Specific applications in the case of the 
Pauli equation and the Rashba and/or Dresselhaus spin-orbit interactions (SOI)
in 2DEGs were analysed in refs.~\cite{Rebei,Jin,Leurs,Medina,Tokatly,BercheEtAl09,DartoraCabrera10,FujitaEtAl2011}.  
{Applications of $SU(2)$ gauge formalism to the case of transport in a superconductor with SO coupling is studied via Green functions formalism in~\cite{Konschelle}.
}

Because the non-Abelian gauge fields components $W^a_i$ are proportional to the electric field, gauge transformations $U$ of the system's Hamiltonian $H$ amount to transformations to a form $UHU^{-1}$ that is unitarily equivalent,  {hence} iso-spectral, but describes in general a different dynamical situation.   {Indeed,  non-Abelian gauge transformations $W^a_i\to W^a_i+\partial_i\alpha^a-\epsilon_{abc}\alpha^b W^c_i$
 contain a term corresponding to an internal rotation. In the case of the SO gauge field, the internal space is spin space and a rotation in spin space as described by the term $\epsilon_{abc}\alpha^b W^c_i$ in the gauge field transformation is precisely what SOI does: it produces spin precession and is explicitly used to manipulate spin orientation in 2DEGs. In the case of Abelian theories, one needs singular gauge transformations to produce a new physical situation by a modification of the curvature~\cite{Kleinert}, here it may be realized even with a global gauge transformation.} 
Nevertheless, there remain classes of gauge transformations that preserve the physics, that have been exploited in several papers in the context of 2DEGs~\cite{LiuZhuZhu,ChenChang,Shikakhwa,Levitov2003,TokatlySherman2010,Adagidelli2012,Gorini2012}. In the first of these papers \cite{LiuZhuZhu} a two-arms spin interferometer with Rashba SOI is reduced to a system with two independent $U(1)$ fields through a $SU(2)$ gauge transformation. In ref.~\cite{ChenChang}, a 2DEG with both Rashba  and Dresselhaus  SO interactions is studied in the Abelian limit of equal  SOI amplitudes. Chen and Chang then use an Abelian gauge transformation to map the system on a free particle model and they recover the phenomenon of Persistent Spin Helix (PSH) first analysed by Bernevig, Orenstein and Zhang~\cite{Bernevig}. {Other authors have already used gauge transformations to simplify SO-related problems. Levitov and Rashba  consider SO coupling in an anisotropic semiconductor quantum dot where the behaviour of electrons is governed by a Schršdinger equation. A gauge transformation was used to turn the SOI into a Zeeman interaction~\cite{Levitov2003}.
In \cite{TokatlySherman2010}, Tokatly and Sherman consider a 2DEG with SO interaction where they analyse diffusive spin dynamics and precessional spin dynamics. A gauge transformation is used to gauge away a pure gauge $SU(2)$ field.
In \cite{Adagidelli2012} the authors study a non relativistic 2DEG with weak inhomogeneous SO coupling to propose a spin transistor which works not only in the ballistic regime, but also in the diffusive regime. Their approach is based on Onsager reciprocity relations,  and they use a local $SU(2)$ gauge transformation to get a spin diagonal structure in the perturbative regime. 
In a similar perspective in \cite{Gorini2012} the authors analyse a 2DEG with SO coupling in relation with Onsager relations to relate the vanishing of spin Hall conductivity when there is only Rashba SO and the finite inverse spin Hall effect. An $SU(2)$ gauge transformation allows them to shift the SO coupling to a Zeeman interaction. 
}

A similar gauge formulation is valid in the relativistic case leading to a theory with a structure 
very similar to the electroweak theory~\cite{Paschos}  (in two space 
dimensions~\cite{Katsnelson}) where ${\cal L}=i\hbar\bar\Psi\gamma^\mu{\nadermat}_\mu\Psi$. Here, $\Psi$ is a two component spinor.
The Dirac $\gamma^\mu$ matrices satisfying Clifford algebra $\{\gamma^\mu,\gamma^\nu\}=2\eta^{\mu\nu} \nbOne$
 (with $\eta^{\mu\nu}=\textmd{Diag}[1,-1,-1]$), are at least $2\times 2$ matrices~\cite{FewBody} and a representation can be given in terms of Pauli matrices
$\gamma^0=\sigma_z$, $\gamma^1=i\sigma_y$, $\gamma^2=-i\sigma_x$. The Dirac equation in this case is a $2\times 2$ matrix equation which, in the absence of an external $U(1)$ field, takes the form
\be
H_D=c{\bm \alpha}\cdot {\bm p} +{\bm \beta} mc^2,
\label{Dirac1}
\ee
where one identifies $\alpha^1=\sigma_x$ and $\alpha^2=\sigma_y$, and $\beta=\sigma_z$ ($\gamma^0=\beta$ and ${\pmb{\gamma}}=\beta{\pmb{\alpha}}$). Like in the non-relativistic case, the non-Abelian gauge field components will follow from the identification of the interactions coupled to the electrons spins through the minimal coupling.  

In contrast to the non-relativistic case, the dispersion relation of massless Dirac particles is linear in momentum so the SOI, can be embedded as a new $SU(2)$ gauge field. As the new concocted gauge fields, built by analogy to a gauge formulation
are also fixed by the physical situation, we can then exploit the whole machinery of gauge transformations in a completely different fashion (not contemplated even for quadratic dispersions): Instead of seeking special classes of transformations that preserve the physics, {\it we determine the gauge transformations that change it} (because material parameters and fields get modified), in order to consider new non-trivial physical situations which, to our knowledge, have not been addressed directly. This is the case of an in-plane electric field (the Rashba interaction corresponds to an out-of-plane electric field), or the case of non uniform (i.e. spatially dependent) gauge parameters. 

The paper is organised as follows. In section 2 we revisit the Hamiltonian of single layer graphene with SOI and show 
how it can be rewritten in terms of a fictitious gauge field. In section 3 we choose a class of gauge transformations
that mute the regular Rashba SOI auxiliary gauge field into a tunable combination of Rashba, Dresselhaus and in plane electric field contributions. Section 4 is devoted to treat the problem of connecting the free graphene without interactions into a spatially dependent SOI. Finally, section 5 addresses the generation of Stark interaction in graphene~\cite{StarkCase}.  As these transformations are done exactly we also obtain the wave functions for a completely new physical situation, thus completely solving the problem with new couplings. The last section summarizes the paper.

\section{The graphene Hamilonian}
The Kane-Mele Hamiltonian~\cite{Kane}, a continuum Hamiltonian derived from a tight binding formulation expanded around the $\vac K$ points, includes both intrinsic and Rashba spin-orbit interactions,
\be
H_{K}=v_F\ \!\bsigma\cdot\vac p
+\DeltaSO\sigma_z
\mats_z
+\lambdaSO(\sigma_x
\mats_y-\sigma_y
\mats_x).\label{eq-HfirstKM}
\ee
$v_F$ is the Fermi velocity, the parameter $\DeltaSO$ measures the strength of the intrinsic SO coupling, and $\lambda_R$ is the Rashba
spin-orbit strength whose source is a Structural Inversion Asymmetry (SIA) and is proportional to an external out-of-plane electric field.
The origin of term springs from a tight-binding approach that couples nearest-neighbour $p_z$ orbitals through paths that
involve $p_z-p_{x,y}$ coupling to the graphene plane, a $V_{sp\sigma}$ hopping integral, and then a Stark $s-p_z$ coupling.
So the $\lambda_R$ involves both the external electric field and the material parameters.

In order to introduce gauge fields, we use the minimal coupling and write equation~(\ref{eq-HfirstKM}) in the form of
$v_F\ 
\bsigma\cdot (\vac p
-\vac\matW^a\ 
\mats_a)$. There is nevertheless a caveat there, since this expression is not Hermitian if $\bsigma$ does not commute with  $\vac p
-\vac\matW^a\ 
\mats_a$. The recipe in that case is  to work with symmetrized expressions.
Once properly symmetrized to generally account for coordinate transformations~\cite{BercheEtAl09}, 
the Hamiltonian can be written as
\be
H_{K}=v_F\ 
\bsigma\cdot (\vac p
-\vac\matW^a\ 
\mats_a)
+\frac 12 v_F[\ \!\vac p
-\vac\matW^a \mats_a,
\bsigma
\ \!],\label{eq-Hfirst}\ee
where the last commutator is  $-i\hbar\left({\pmb{\nabla}}\cdot\bsigma\right)$.  
When both Rashba and intrinsic SOI are present, inspection of Eq.~(\ref{eq-HfirstKM}) shows  that the gauge fields take the form 
\begin{eqnarray}
\vac W^1=(\lambdaSO/ v_F)\vac u_y,~ &\vac W^2&=-(\lambdaSO/v_F)\vac u_x,\notag \\
 \vac W^3&=&-(\DeltaSO/v_F)\vac u_z,
\end{eqnarray}
where $\vac u_i$ are unit vector in the $i$-th cartesian coordinate. $\lambdaSO$ and $\DeltaSO$ both have the dimensions of inverse time
 and $W^a_i$ those of an inverse length. 

 {We now reemphasise that the present gauge fields depend on physical given quantities so that there is no gauge freedom\cite{Medina,BercheEtAl12} (an introductory review can be found in Ref.~\cite{EJP}). We take advantage of this peculiarity of the gauge structure of the spin-orbit interaction to find a gauge transformation that connects {\it two physically distinct problems}. This we develop in the next section. }

\section{Transforming spin-orbit couplings on a graphene sheet \label{sec3}}

We will now contemplate the case of spin-orbit interaction which  can be described in terms of
non-Abelian gauge theory.  Our aim is to generate an interaction or modify an interaction already present. 
We can transform the 
 dynamical momentum ${\pmb{\pi}}=\vac p
 -\vac W^a\mats_a$ which follows from  the minimal coupling prescription, 
 under a non-Abelian gauge transformation
parameterized by the set of three
functions $\alpha^a$ (see e.g. Ref.~\cite{EJP}) in  $U_s=e^{\frac{i}{\hbar}\frac{\alpha^a}2\mats_a}$ according to 
\bey
\vac p-{\vac W'}^a\mats_a&=&
 {\pmb{\pi}}'=U_s\ {\pmb{\pi}}\ U_s^{-1}
\nonumber\\
&=&\vac p
-\left(\vac W^a+{\pmb{\nabla}}\alpha^a-\epsilon_{abc}\alpha^b\vac W^c\right)\ \mats_a
\eey
up to $O(|\alpha|^2)$ terms, where $|\alpha|=\sqrt{\alpha^a\alpha_a}$. The transformation of the gauge potential thus assumes the form
 ${\vac W'}^a=\vac W^a+{\pmb{\nabla}}\alpha^a-\epsilon_{abc}\alpha^b\vac W^c$.  {Even a set of constant gauge transformation parameters $\alpha^a$ will in general create new interactions ${\vac W'}^a$ from original ones ${\vac W}^a$ through the last term of the equation. On the other hand, if one starts at the outset from the free particle case, non uniform $\alpha^a(\vac r)$ are required to generate 
 non trivial effects.}

Let us consider a graphene sheet and 
contemplate an initial problem with a non zero spin-orbit interaction, e.g. Rashba SOI due to the presence of an electric field $E_\perp$ perpendicular to the sheet (here treated in Cartesian coordinates at one of the Dirac points),
\be H_{\rm R}=v_F\ \bsigma\cdot\vac p
+
\lambdaSO(\sigma_x
\mats_y-\sigma_y
\mats_x).\label{eq-HR}\ee 
 Acting on a spinor in the form of a two-dimensional plane wave multiplied by 4-component space-independent amplitudes, 
 the Hamiltonian of Eq.~(\ref{eq-HR}) leads to the secular equation 
 $(E/\hbar v_F)^4-2(E/\hbar v_F)^2(|\vac k|^2+2(\lambdaSO/v_F)^2)+|\vac k|^4=0$. The eigenvalues follow~\cite{StauberSchliemann09} 
 \be 
E_{\kappa,\delta}=\kappa\hbar v_F\left (\delta\lambdaSO/v_F+\sqrt{|\vac k|^2+(\lambdaSO/v_F)^2}\right),
 \label{Eq9}
 \ee where  $\kappa=\pm 1$ is the particle-hole index and $\delta=\pm 1$ labels the SO branches.

We will investigate the transformation  $U_s=\exp\left (\frac{i}{\hbar}\frac{\alpha^a}{2}\mats_a\right) $, of this Hamiltonian
under a set of {\it constant} gauge parameters $\alpha^a$. In this case the gauge operations are pure rotations of the spin degree of freedom
so one can generate very simple interaction transformations.  The free particle term is invariant and we have to consider the transformation of terms like
\bey
  \mats_b\to \mats'_b&=&  \exp\left (\frac{i}{\hbar}\frac{\alpha^a}2\mats_a\right ) \mats_b\exp\left (-\frac{i}{\hbar}\frac{\alpha^c}2\mats_c\right )\nonumber\\
 \phantom{\mats_b\to \mats'_b}& =&\cos^2{\textstyle\frac{|\alpha|}{2}}\mats_b-\frac {2}{|\alpha|}\cos{\textstyle\frac{|\alpha|}{ 2}}\sin{\textstyle\frac{|\alpha|}{2}}\epsilon_{abc}\alpha^a\mats_c
 \nonumber\\
 \phantom{\mats_b\to \mats'_b}&&
 +\frac 1{(\alpha)^2}\sin^2\left ({\textstyle\frac{|\alpha|}{2}}\right )\alpha^a\alpha^c(\delta_{bc}\mats_a-\delta_{ac}\mats_b+\delta_{ab}\mats_c).\nonumber\\
 \eey
 A case which has enough generality for our purpose is given via the choice $\alpha^1=\alpha\cos\phi$, $\alpha^2=\alpha\sin\phi$, $\alpha^3=0$, $\phi={\rm const.}$ A rich variety of spin-orbit interactions are generated, e.g. when we fix $\phi=\pi/4$,
 \bey
 &&\lambdaSO(\sigma_x%\otimes
  \mats_y-\sigma_y%\otimes
  \mats_x)\to \lambdaSO\cos^2\frac{|\alpha|}{2}(\sigma_x%\otimes
  \mats_y-\sigma_y%\otimes
  \mats_x) \nonumber\\ &&\quad
 +\lambdaSO\sin^2\frac{|\alpha|}{2}(\sigma_x%\otimes
 \mats_x-\sigma_y%\otimes
 \mats_y)+\frac\lambdaSO{\sqrt 2}\sin\alpha(\sigma_x%\otimes
 \mats_z-\sigma_y%\otimes
 \mats_z),\nonumber\\
 \eey i.e. one generates a
 Dresselhaus interaction with amplitude $\lambda'_{\rm D} = \lambdaSO\sin^2\alpha/2$ and a Pauli spin-orbit interaction which comprises a term due to a perpendicular electric field (ordinary Rashba term with amplitude $\lambdaSO' = \lambdaSO\cos^2\alpha/2$) plus another term due to an in plane electric field with amplitude $\lambda'_{\parallel}=\frac\lambdaSO{\sqrt 2}\sin\alpha$. 
 
Interesting limiting cases  are  found when the amplitude $\alpha$ is fixed: 

i) $\alpha=\pi$, which amounts to a rotation by this angle around the $x=y$ axis, where the transformed problem exhibits {\it a pure Dresselhaus interaction}, and 

ii) $\alpha=\pi/2$ where Rashba and Dresselhaus have equal amplitudes and the in-plane Pauli interaction is also present in the transformed problem. Case i) interestingly shows that Rashba and Dresselhaus interactions are related through a gauge transformation. This case has been addressed in the literature\cite{LossR+D}. 

For case ii) the transformed Hamiltonian reads as
 \bey 
   H'_{\rm R}&=&v_F\ \bsigma\cdot\vac p%\otimes\nbOne_s
+\lambda'_{\rm R}(\sigma_x%\otimes
\mats_y-\sigma_y%\otimes
\mats_x)\nonumber\\
&&
+\lambda'_{\rm D}(\sigma_x%\otimes
\mats_x-\sigma_y%\otimes
\mats_y)+\lambda'_{\parallel} (\sigma_x%\otimes
\mats_z-\sigma_y%\otimes
\mats_z)\nonumber\\
\label{eq-HRprime}\eey
The corresponding (un-normalized) eigenvectors of the initial Hamiltonian are given by
 \be
 \Psi(\vac r)=e^{i\vac k\vac r}\begin{pmatrix}
 ie^{-i\phi_{\vac k}}\cos\frac{\theta_{\kappa,\delta}}2\\
 \kappa\delta\sin \frac{\theta_{\kappa,\delta}}2\\
 i\sin\frac{\theta_{\kappa,\delta}}2\\
 \kappa\delta e^{i\phi_{\vac k}}\cos\frac{\theta_{\kappa,\delta}}2
 \end{pmatrix},\vspace{0.5cm}
 \ee 
 with $\tan\phi_{\vac k}=k_y/k_x$ and $\tan\theta_{\kappa,\delta}=E_{\kappa,\delta}/\hbar v_F|\vac k|$, $\theta_{\kappa,\delta}\in[-\pi,\pi]$.

Acting on these eigenstates, with the unitary gauge transformation
 \bey
&& \nbOne_\sigma\otimes\ U_s  %\Psi(\vac r)
 =\nonumber\\
 &&\ =\begin{pmatrix}
 \cos{\textstyle\frac{|\alpha|}{2}} & 0 & {\textstyle\frac {1+i}{\sqrt 2}}\sin {\textstyle\frac{|\alpha|}{2}} & 0 \\
0 & \cos{\textstyle\frac{|\alpha|}{2}} & 0 & {\textstyle\frac {1+i}{\sqrt 2}}\sin {\textstyle\frac{|\alpha|}{2}} \\
- {\textstyle\frac {1-i}{\sqrt 2}}\sin {\textstyle\frac{|\alpha|}{2}} &0& \cos{\textstyle\frac{|\alpha|}{2}} & 0  \\
0&- {\textstyle\frac {1-i}{\sqrt 2}}\sin {\textstyle\frac{|\alpha|}{2}} &0& \cos{\textstyle\frac{|\alpha|}{2}} 
 \end{pmatrix} %\Psi(\vac r)
 \nonumber\\
 \eey
 one obtains  the eigenstates of Hamiltonian~(\ref{eq-HRprime}). 
 \bey
&&\Psi'(\vac r) =\nonumber\\
&&\ =
e^{i\vac k\vac r}
 \begin{pmatrix}
 ie^{-i\phi_{\vac k}}\cos\frac{|\alpha|}{2}\cos\frac{\theta_{\kappa,\delta}}2
 +i\frac{1+i}{\sqrt 2}\sin\frac{|\alpha|}{2}\sin\frac{\theta_{\kappa,\delta}}2
 \\
 \kappa\delta\cos\frac{|\alpha|}{2}\sin \frac{\theta_{\kappa,\delta}}2
 +\kappa\delta\frac{1+i}{\sqrt 2} e^{i\phi_{\vac k}}\sin\frac  \alpha 2\cos\frac{\theta_{\kappa,\delta}}2\\
-i\frac{1-i}{\sqrt 2} e^{-i\phi_{\vac k}}\sin\frac{|\alpha|}{2}\cos\frac{\theta_{\kappa,\delta}}2
+i\cos\frac{|\alpha|}{2}\sin\frac{\theta_{\kappa,\delta}}2\\
 -\kappa\delta\frac{1-i}{\sqrt 2}\sin\frac{|\alpha|}{2}\sin \frac{\theta_{\kappa,\delta}}2
+ \kappa\delta e^{i\phi_{\vac k}}\cos\frac{|\alpha|}{2}\cos\frac{\theta_{\kappa,\delta}}2
 \end{pmatrix}\nonumber\\
 \eey
The eigenvalues are the same as those of Eq.\ref{Eq9}.  As the physical fields and the material parameters generating the couplings in the above Hamiltonian are different from those of the original problem, we have then fully solved a new physical situation.

There is nevertheless a physical effect that we have not taken into account. When there is an in plane electric field (as the one which enters the spin-orbit interaction in the transformed model), one has to add the Stark contribution to the potential energy, which, although spin-independent, has a spatial dependence which modifies the physical problem under study. This example allows us to illustrate another 
route to generate spin-orbit interactions from the gauge transformation. 

\section{Spatially dependent transformations\label{sec4}}
We now contemplate  a graphene sheet and start from the free particle problem (at one Dirac point) in Cartesian coordinates,
 \begin{equation}
 H_0=v_F\ \bsigma\cdot\vac p
 =-i\hbar v_F(\sigma_x\partial_x+\sigma_y\partial_y),\label{Eq15}
 \end{equation}
and assume a gauge transformation $U_s=\exp(\frac{i}{\hbar}\frac{\alpha^a}2\mats_a) $ with a  space-dependent set of $\alpha^a(\vac r)$. {The idea is here to generate local SOI by rendering local the gauge structure of the theory.} The gauge transformation of the first term {in Eq.~(\ref{Eq15})}
reads as 
\be(i\partial_x)\to \exp(\frac{i}{\hbar}\frac{\alpha^a}2\mats_a) (i\partial_x)\exp(-\frac{i}{\hbar}\frac{\alpha^c}2\mats_c).\label{Eq17}
\ee The transformed operator {at the r.h.s. of Eq.~(\ref{Eq17})}, being a $2\times 2$ matrix,  {it} can be rewritten in terms of Pauli matrices (in real spin space):
\bey
  i\partial_x
&\to& i\partial_x+\frac 12(\partial_x\alpha)\frac{\alpha^b\mats_b}{\hbar\alpha}
\nonumber\\
&&
-\frac 1\alpha\sin\frac{|\alpha|}{2} \left[(\partial_x\alpha^c) -\frac{\alpha^c (\partial_x\alpha)}{\alpha}\right]
\times
\nonumber\\
&&\times \left[\cos\frac{|\alpha|}{2}\frac{\mats_c}{\hbar}+i\frac{\alpha^a}\alpha\sin\frac{|\alpha|}{2} (\delta_{ac}\nbOne_s-i\epsilon_{abc}\frac{\mats_b}{\hbar})\right]\label{Eq18}
\eey  
This expression is very involved, but it clearly shows how diverse forms of spin-orbit interactions may be realized after the tensor product between
the pseudo-spin matrices. As an example, let us consider a space-dependent gauge parameter {consisting in only one spin space component $\alpha^3$ which depends on the position in the graphene sheet,} $\alpha^a\equiv\alpha^3(x,y)$. Then, the free particle Hamiltonian 
$v_F\ \bsigma\cdot\vac p%\otimes\nbOne_s
$ {in Eq.~(\ref{Eq15}) becomes} 
\bey H_0'&=&v_F\ \bsigma\cdot\vac p%\otimes\nbOne_s
-
\frac{1}{2}(\partial_x\alpha^3 \sigma_x%\otimes
\mats_z + \partial_y\alpha^3 \sigma_y%\otimes
\mats_z )
\nonumber\\ &=& 
v_F\ \bsigma\cdot
(\vac p%\otimes\nbOne_s
-\vac W^a(x,y)\mats_a) .\label{Eq19}
\eey 
The complete solution {for  the ``in-plane field'' Hamiltonian~(\ref{Eq19})} can {now be obtained via the unitary spatially dependent gauge transformation 
$U_s=\exp(\frac{i}{\hbar}\frac{\alpha^3(x,y)}2\mats_z) $ applied to the   free problem (\ref{Eq15}). An example is treated in the next section.}

\section{The Stark coupling\label{sec5}}

The transformation performed in section~\ref{sec3}, generates an 
in-plane electric field $\vac{\cal E}=({\cal E}_x,{\cal E}_y)$. Then, the associated contribution to the potential energy must be taken into account. For simplicity, we consider the case of a uniform electric
field. The Stark potential energy $-e\vac {\cal E}\cdot\vac r$,  in terms of a tight-binding Hamiltonian\cite{Pastawski}, and then expanded in the vicinity of the Dirac point in the continuum limit, reads as $H_{\rm stark}=\omega_{xx}\sigma_x p_x + \omega_{xy}\sigma_x p_y + \omega_{yx}\sigma_y p_x + \omega_{yy}\sigma_y p_y$ (up to a quantity proportional to the identity that we can absorb in the redefinition of the zero of the energies). The coefficients are given in terms of the hopping elements along the bonds of the lattice that are now anisotropic and are labelled as $\gamma_i$:
$\omega_{xx}=\frac {\sqrt 3a}{4\hbar}\tau(\gamma_2+\gamma_3)$, 
$\omega_{xy}=\frac {\sqrt 3a}{2\hbar}\tau(\gamma_2-\gamma_3)$,
$\omega_{yx}=-\frac {a}{4\hbar}(\gamma_2-\gamma_3)$,
$\omega_{yy}=-\frac a\hbar(\gamma_1+\frac 12(\gamma_2+\gamma_3))$. Only the form of the Hamiltonian is important here.

When the Stark contribution is added to the ordinary kinetic energy, one is thus led to an anisotropic kinetic energy which reflects the drift induced by the field: $v_{xx}\sigma_x p_x + v_{xy}\sigma_x p_y + v_{yx}\sigma_y p_x + v_{yy}\sigma_y p_y$ with
$v_{ii}=v_F+\omega_{ii}$ and $v_{ij}=\omega_{ij}$. An interesting simplification occurs when the in-plane electric field is along the $\gamma_1$ bond,  since then $\gamma_2=\gamma_3$ and $\omega_{xy}=\omega_{yx}=0$. If now one considers the consistent Hamiltonian comprising the ordinary kinetic energy and the Stark and spin-orbit contributions associated to an in-plane electric field $\vac {\cal E}={\cal E}{\vac u}_y$, one has
\be
H=v_{xx}\ \sigma_x p_x+v_{yy}\ \sigma_y p_y
+
\frac{e v_F}{4mc^2}(\vac{\cal E}\times\bsigma)_z\mats_z
\label{eq32}
\ee
Devising a transformation that eliminates the space-de\-pen\-dent SOI term, we look for one that is the inverse 
to the one that {\it created} the interaction i.e.
$U_s=\exp\left (-\frac{i}{\hbar}\frac{\alpha^3}2\mats_z\right ) $.  
{Using Eq.~(\ref{Eq18}), one gets $p_{x,y}\to p_{x,y}-\frac 12(\partial_{x,y}\alpha^3)s_z$ and}
the transformed Hamiltonian is now
  \bey
   {H'=}U_sHU_s^{-1} &=& v_{xx}\ \sigma_x p_x+v_{yy}\ \sigma_y p_y
%+\Delta_{\rm SO}\sigma_z\mats_z
\nonumber\\
&&-\left(\frac 12v_{xx}\partial_x\alpha^3+\frac{e v_F {\cal E}}{4mc^2}\right)\sigma_x 
\mats_z\nonumber\\
&&-\left(\frac 12v_{yy}\partial_y\alpha^3\right)\sigma_y
 \mats_z.
  \eey
The cancellation of the spin-dependent terms is achieved by choosing $\alpha^3(\vac r)=-\frac{ev_F{\cal E}}{2mc^2v_{xx}}x$ and the Hamiltonian thus becomes
\be
H'=v_{xx}\ \sigma_x p_x+v_{yy}\ \sigma_y p_y.\label{eq22}
\ee
Its eigenvalues and eigenvectors are easily obtained,
\be
\Psi'(\vac r)=e^{i(k_xx+k_yy)}\begin{pmatrix}
F_\uparrow\\F_\downarrow\\G_\uparrow\\G_\downarrow
\end{pmatrix},
\ee
\be E'=\pm\sqrt{(\hbar v_{xx}k_x)^2+(\hbar v_{yy}k_y)^2},\ee
and those of the original problem (\ref{eq32}) then follow via the inverse gauge transformation
\be\Psi(\vac r)=U_s^{-1}\Psi'(\vac r)
=e^{i(k_xx+k_yy)}\begin{pmatrix}
F_\uparrow e^{i\frac{e{\cal E}v_F}{4mc^2v_{xx}}x\phantom-}\\
F_\downarrow e^{-i\frac{e{\cal E}v_F}{4mc^2v_{xx}}x}\\
G_\uparrow e^{i\frac{e{\cal E}v_F}{4mc^2v_{xx}}x\phantom-}\\
G_\downarrow e^{-i\frac{e{\cal E}v_F}{4mc^2v_{xx}}x}
\end{pmatrix}.
\ee

The physical properties being associated to different physical situations, they may differ in the two gauges. 
Consider the example of the current densities. The charge current density in a given eigenstate $|\Psi\rangle$ 
may be defined by the matrix element $\vac j_q = e\langle \Psi |v_F~\pmb{\alpha}%\otimes\nbOne_s
|\Psi\rangle$ where $v_F\ \pmb{\alpha}%\otimes\nbOne_s
=(v_F)^{-1}(v_{xx}\sigma_x{\vac u}_x+v_{yy}\sigma_y{\vac u}_y)%\otimes\nbOne_s
$ is the velocity operator.
This charge current density is invariant under the gauge transformation, since in the primed problem one has
\be\vac {j'}_q = ev_F\langle \Psi |%(\nbOne_\sigma\otimes\ 
U_s^{-1} (\sigma_x{\vac u}_x+\sigma_y{\vac u}_y)%\otimes\nbOne_s (\nbOne_\sigma\otimes\ 
U_s|\Psi\rangle=\vac j_q\ee {because the operator $v_F\ \!\pmb{\alpha}$  in sandwich is spin-independent.} 
  {This is not a surprising result, since as we have noticed the non-Abelian gauge transformation produces a spin rotation and hence does not touch physical properties which are independent of the spin state.}
The situation is different with the spin current density,  
$\vac j_s^a = \langle \Psi |v_F\ \pmb{\alpha}\mats_a|\Psi\rangle$, since then $U_s$ does not commute with $\mats_a$   {and the resulting spin current is depending on the gauge. {Like in the case of the charge current, let us form the spin current in the primed problem in terms of the original eigenstates,
\be\vac {j'}_s^a = \langle \Psi |
U_s^{-1}v_F\ \pmb{\alpha}\mats_a
U_s|\Psi\rangle\ee
where $U_s^{-1}\mats_aU_s \to \cos^2\frac{|\alpha^3|}{2}\mats_a+\sin^2\frac{|\alpha^3|}{2}(2\mats_z\delta_{a3}-\mats_a)$, i.e. $\mats_z\to \mats_z$ because the gauge transformation consists in a rotation around the $z-$axis, while $\mats_{x,y}\to\cos^2\frac{|\alpha^3|}{2}\mats_{x,y}-\sin^2\frac{|\alpha^3|}{2}\mats_{z}$. An example of a spin-dependent physical property which is not the same in the two problems considered is 
\be \vac j_s^x=\langle\Psi| v_F\ \!\pmb{\alpha}\ \!\mats_x |\Psi\rangle\ee
with Hamiltonian~(\ref{eq32}) and 
\be \vac {j'}_s^x=\langle\Psi| v_F\ \!\pmb{\alpha}\ \!\left(\cos^2\frac{|\alpha^3|}{2}\mats_{x}-\sin^2\frac{|\alpha^3|}{2}\mats_{z}\right) |\Psi\rangle\ee
with Hamiltonian~(\ref{eq22}). 
}

This is a known result in non-Abelian gauge theory that the conserved currents are gauge dependent {(gauge covariant)}~\cite{Hughes}}. 
{The spin current furthermore constitues only part of the conserved current and, alone, is not conserved due to the fact that the spin interacts with the non-Abelian gauge fields (i.e. here the electric field), see e.g. Ref.~\cite{Maekawa}.}

\section{Summary and Conclusions\label{sec4}}

We have analysed $SU(2)$ gauge transformations in graphene close to the $\vac K$ points.  {As unitary transformations they preserve the eigenvalues, but due to the electric field dependence in the auxiliary gauge potentials, such transformations convert from one physical model to another.} Here we contemplated the connection between the pure Rashba coupling to the pure Dresselhaus coupling and also between the former and a Dresselhaus coupling plus an in-plane electric field. We also showed how one could generate SOI resulting from
in-plane electric fields from the free graphene Hamiltonian thru a spatially dependent gauge transformation. The strengths of the couplings were proportional to the gradient of the space dependent gauge parameter. Physically, the in-plane field effect can also be 
realised in practice for stretched graphene. 

We finally considered a series of tailored gauge transformations in order to address the consistent Hamiltonian for an in-plane field
plus its corresponding Stark coupling. in this case, we transformed the problem into an anisotropic  Dirac cone problem with intrinsic SOI. For this problem, the eigenfunctions can be fully derived and transformed back to obtain the corresponding ones for the original Hamiltonian.

We must note that the gauge transformation fixes the relative values of the interactions generated as a function of the
gauge transformation chosen and the original parameters. If there is a threshold for a phase transition as a function of the parameter
values, the transformation will in general not address it.

We have stressed that the $SU(2)$ gauge transformation applied to the Dirac Hamiltonian, for the three problems considered before, now corresponds to different physical situations, since the non-Abelian gauge field components are bound to real external fields and material parameters.  In this manner, we have shown that the corresponding Hamiltonian can be conveniently addressed by means of the aforementioned applied gauge transformation. The transformation however  keeps the eigenenergies unchanged, so the modification of eigenvectors {and of spin-dependent operators after gauge transformation now allows for different physical expectation values regarding for example spin persistent currents. This is connected to the covariance property of $SU(2)$ gauge transformations which amount to rotations in internal space which here is the ordinary space coupled to spin degrees of freedom.}

\section*{Acknowledgment} NB thanks the Coll\`ege Doctoral franco-allemand 02-07 ``Physics of Complex Systems'' for financial support, 
EM and BB are respectively grateful to the University of Lorraine and to IVIC for support. They also thank the CNRS and FONACIT 
for support through the ``PICS'' programme {\it Spin transport and spin manipulations in condensed matter: polarization, spin currents and entanglement}. AL thanks the Statistical Physics Group in Nancy for its support for short visits. 

{\section*{Author contribution statement} All authors have equally contributed to this work.}

\end{document}